\documentclass[twocolumn]{revtex4-1}
\usepackage[T2A]{fontenc}
\usepackage[cp1251]{inputenc}
\usepackage[english]{babel}
\usepackage{amsmath}
\usepackage{amsfonts}
\usepackage{amssymb}
\usepackage{makeidx}
\usepackage[colorlinks=true]{hyperref}
\usepackage[pdftex]{graphicx}

\usepackage{color}

\newcommand{\sech}{\mathop{\rm sech}}

\newcommand{\e}{{\mathop{\rm e}}}
\renewcommand{\d}{d}

\begin{document}

\title{Spin noise of localized electrons interacting with optically cooled nuclei}

\author{D.~S.~Smirnov}
\email[Electronic address: ]{smirnov@mail.ioffe.ru}

\affiliation{Ioffe Physical-Technical Institute of the RAS, 194021, St.-Petersburg, Russia}

\begin{abstract}

A microscopic theory of spin fluctuations of localized electrons interacting with optically cooled nuclear spin bath has been developed. Since nuclear spin temperature may stay low enough for macroscopically long time, the nuclear spin system becomes very sensitive to an external magnetic field.
This strongly affects electron spin noise spectrum. It has been shown that in the case of weak fields/relatively high nuclear spin temperature, a small degree of nuclear spin polarization affect the electron spin fluctuations in the same way as an additional external magnetic field. By contrast, the high degree of nuclear polarization realized in relatively strong magnetic field and low nuclear spin temperature leads to a suppression of hyperfine field fluctuations and to a dramatic narrowing of precession-induced peak in the spin noise spectrum. The experimental possibilities of nuclear spin system investigation by means of spin noise spectroscopy are discussed.

\end{abstract}

\maketitle

\section{Introduction}

In the past few years spin noise spectroscopy has emerged as a new experimental tool for spin dynamics 
investigation~\cite{Zapasskii:13,Oestreich-review,Mueller2010}. This method was originally proposed for atomic gases~\cite{aleksandrov81}, and 
than transferred to semiconductors~\cite{Oestreich_noise}. The spin noise is usually detected via fluctuations of spin Kerr, Faraday or 
ellipticity signals~\cite{noise-trions}. The autocorrelation function of the spin signals contains information about precession frequencies, spin relaxation 
times and other parameters of spin dynamics. For the light propagating along $z$-axis the Fourier transform of this correlator is proportional to the spectrum 
of omnipresent spin fluctuations $(s_z^2)_\omega$.

Spin noise spectroscopy being a nearly perturbation-free technique is especially useful for the studies of slow spin 
dynamics in equilibrium or close-to-equilibrium conditions~\cite{muller-Wells,crooker2010}. A vast variety of systems was studied by 
this method, among them are bulk semiconductors~\cite{PhysRevB.79.035208,Romer2010}, multiple and single quantum-well 
structures~\cite{muller-Wells,noise-trions}, single quantum dots (QDs)~\cite{singleHole} and quantum-dot ensembles~\cite{crooker2010}. 
Moreover it has been found that the spin noise spectroscopy in non-equilibrium conditions can provide complementary information to the time-resolved 
photoluminescence spectroscopy or pump-probe techniques~\cite{Sinitsyn,Marina-noise,noise-excitons,Mollow-noise}.

Excitation of ensemble of localized electrons by circularly polarized light leads not only to carrier spin orientation, but also to dynamical polarization of host lattice
nuclei~\cite{OptOr}. After switching off the pump the electron spin system relatively fast returns to the equilibrium within 
several hundreds of nanoseconds in bulk semiconductors~\cite{Dzhioev02,PhysRevB.85.195313}.
The dynamics of nuclear spin system is more complex: Usually, on the microsecond time scale the nuclear system reaches its equilibrium with the effective nuclear spin temperature, while the relaxation of nuclear spin temperature or longitudinal relaxation of nuclear spins takes place on the macroscopic time scale, from seconds to days~\cite{Urbaszek}. During that time the electron and nuclear subsystems can be characterized by the spin temperatures differing by 
several orders of magnitude~\cite{fleisher1976optical}. Application of magnetic field to such system can weakly affect the electron spin dynamics, but induce significant nuclear polarization~\cite{OptOr,dyakonov1975}. In this paper we study the electron spin fluctuations in the steady-state but non-equilibrium system consisting of ``cold'' nuclear spin bath and ``hot'' localized electrons in the presence of magnetic field. We show that the spin noise spectra reflect the distribution function of hyperfine field and 
that the nuclear spin polarization 
drastically modifies the electron spectra.


The paper is organized as follows. In Sec.~\ref{sec:model} we obtain the general expression for the spin noise spectra of localized electrons interacting with the cooled nuclei. 
Secs.~\ref{sec:faraday} and~\ref{sec:voigt} are devoted to the spin noise spectra in the Faraday and Voigt geometries, respectively. In Sec.~\ref{sec:I} we describe the spectra for different magnitudes of the nuclear spins. Finally, in Sec.~\ref{sec:conclude} we conclude with the discussion of possible experimental realizations.

\section{Model}
\label{sec:model}

We consider an ensemble of isolated electrons localized on donors or in QDs. Each electron effectively interacts with large amount of nuclei $N\gg1$, each having the spin~$I$. In typical III-V compounds such as GaAs, InP or InAs each nucleus possesses nonzero spin, and $N\sim10^{5}$, but in the solids with very few isotopes carrying a nuclear spin (for example, ZnO or CdSe) $N$ can be of the order of $100$. The hyperfine interaction results in random effective magnetic fields acting on electron spin. The quantum-mechanical average of the given electron spin, $\mathbf{s}$, obeys the
Bloch equation
\begin{equation}
 \frac{\d\mathbf{s}}{\d t}=(\mathbf{\Omega}_N+\mathbf{\Omega}_B)\times\mathbf{s}-\frac{\mathbf{s}}{\tau_s}.
 \label{Bloch}
\end{equation} 
Here $\tau_s$ is the phenomenological electron spin relaxation time caused by, e.g., electron-phonon interaction~\cite{PhysRevB.64.125316}, $\mathbf{\Omega}_N$ and $\mathbf{\Omega}_B$ are the Larmor precession frequencies related to the hyperfine interaction and external magnetic field $\mathbf{B}$. In fact $\tau_s$ can depend on magnetic field and nuclear spin polarization but these effects are neglected in this paper for the sake of simplicity. The Knight field nuclear spin dynamics takes place on the time scale $\sim\sqrt{N}/\Omega_N$~\cite{merkulov02,OptOr}, and in what follows we restrict ourself to the model of ``frozen'' nuclear fluctuation by assuming that $\sqrt{N}/\Omega_N\gg\tau_s$~\cite{noise-CPT}.

For the given localized electron the Larmor precession frequency in the random hyperfine field can be expressed as
\begin{equation}
 \mathbf{\Omega}_N=\sum_{k=1}^NA_k \mathbf{I}_k,
\end{equation} 
where $\mathbf{I}_k$ are the spins of nuclei interacting with the electron and $\hbar A_k$ are the corresponding hyperfine constants.
In thermal equilibrium and in the absence of external magnetic field the nuclear spins are randomly oriented giving rise to mean square fluctuation~\cite{merkulov02}
\begin{equation}
\left\langle\mathbf{\Omega}_N^2\right\rangle=I(I+1)\sum_{k=1}^NA_k^2\equiv\frac{3}{2}\delta^2,
 \label{delta}
\end{equation} 
where the angular brackets denote the averaging over the time and/or over the ensemble~\footnote{In QD ensembles an additional averaging over the contributions of different QDs has to be performed.} and we have introduced parameter $\delta$ describing the dispersion of the hyperfine field. We assume that the cartesian components of the hyperfine field are independent of each other, thus in the absence of external field and optical pumping the nuclear field is described by the normal distribution
\begin{equation}
\mathcal F_0(\mathbf{\Omega}_N)=f(\Omega_{N,x},\delta)f(\Omega_{N,y},\delta)f(\Omega_{N,z},\delta), 
\label{F0}
\end{equation} 
where 
\begin{equation}
 f(x,\sigma)=\frac{1}{\sigma\sqrt{\pi}}\e^{-x^2/\sigma^2}.
 \label{distrib}
\end{equation} 
is the Gaussian distribution function.

The spin noise spectrum of single electron can be calculated using the method of Langevin random forces developed in~\cite{NoiseGlazov}, or Heisenberg representation for spin operators~\cite{noise-exchange-eng}, or the quantum regression theorem~\cite{Glazov_Wire} and then it should be averaged over the distribution function of nuclear fields $\mathcal F(\mathbf{\Omega}_N)$. The final result for the spin noise spectrum reads~\cite{NoiseGlazov}
\begin{multline}
 (s_z^2)_\omega=\frac{\pi}{2}\int\d\mathbf{\Omega}\mathcal F(\mathbf{\Omega}-\mathbf{\Omega}_B)\left\lbrace\frac{\Omega_z^2}{\Omega^2}\Delta(\omega)+\right. \\
 \left.\frac{\Omega_x^2+\Omega_y^2}{2\Omega^2}\left[\Delta(\omega-\Omega)+\Delta(\omega+\Omega)\right]\right\rbrace,
 \label{general}
\end{multline} 
where
\begin{equation}
 \Delta(x)=\frac{1}{\pi}\frac{\tau_s}{1+(x\tau_s)^2}
\end{equation} 
is the broadened $\delta$-function. The above general expression is valid whenever the model of ``frozen'' nuclear fluctuation can be applied.
Further on the basis of Eq.~\eqref{general} we will analyze the spin noise in equilibrium conditions, in Faraday and Voigt geometries.

First of all, in the equilibrium conditions one has to substitute the distribution function $\mathcal F_0(\mathbf{\Omega}_N)$, Eq.~\eqref{F0}, into Eq.~\eqref{general} to obtain~\cite{NoiseGlazov} \begin{multline}
\label{steady}
 (s_z^2)_\omega=
 \frac{\pi}{6}\left\lbrace\Delta(\omega)+\right. \\
 \left.\int_0^\infty\d\Omega F(\Omega)\left[\Delta(\omega-\Omega)+\Delta(\omega+\Omega)\right]\right\rbrace,
\end{multline} 
where $F(\Omega)=4(\Omega^2/\delta^2)f(\Omega,\delta)$ is the distribution function of the hyperfine field absolute value. In what follows we will accept the realistic assumption $\tau_s\delta\gg 1$, which allows one to replace the broadened $\delta$-function in the second term of Eq.~\eqref{steady} by a real $\delta$-function and obtain
\begin{equation}
(s_z^2)_\omega=\frac{\pi}{6}\left[\Delta(\omega)+F(\omega)\right].
\label{steady-limit}
\end{equation} 
Thus the function $F(\omega)$ defines the shape of the spectrum at frequencies $\omega>1/\tau_s$.

The dynamical polarization pushes nuclei out of equilibrium, and at the timescale exceeding $\sim10^{-4}$ s the nuclear spin bath can be described by effective nuclear spin temperature $T_N$~\cite{OptOr}. This temperature can be both positive and negative and as low as a few $\mu$K~\cite{kalevich_82a,dyakonov1975}. The application of magnetic field $B$ exceeding local nuclear fields (few Gauss) to the cold nuclear spin bath cause significant spin orientation~\cite{OptOr,dyakonov1975} and dramatically modifies the spin noise spectra through the distribution function of the hyperfine field. In case of complete nuclear spin polarization the maximum Larmor precession frequency caused by hyperfine field is given by
\begin{equation}
 \Omega_0=I \sum_{k=1}^N A_k.
 \label{omega0}
\end{equation} 
The general expression for the average Larmor frequency $\overline{\Omega}$ for $I=1/2$ nuclei reads
\begin{equation}
 \overline{\Omega}=\Omega_0\tanh\frac{\hbar\gamma B}{2kT_N}\equiv\Omega_0\tanh\frac{B}{B_T}.
\label{P}
\end{equation} 
Here $\gamma=\mu_n g_n/\hbar$ is the nuclear gyromagnetic ratio, $\mu_n$ is the nuclear magneton, $g_n$ is the nuclear $g$-factor and, for the sake of convenience, we have introduced the characteristic magnetic field related to nuclear spin temperature $B_T=2kT_N/(\hbar\gamma)$. The arbitrary magnitude of nuclear spins $I$ will be considered in Sec.~\ref{sec:I}. 
The average Larmor frequency $\overline{\Omega}$ can be comparable and even exceed the precession frequency in the external field $\Omega_B$ for $kT_N\ll\sqrt{N}\hbar\delta\left|\mu_ng_n/(\mu_Bg_e)\right|$, where $g_e$ is the electron g factor and $\mu_B$ is the Bohr magneton. Provided the nuclear spins are uncorrelated the distribution of nuclei-induced Larmor precession frequency is Gaussian and given by~\cite{schulten}
\[
 \mathcal F(\mathbf{\Omega}_N)=f(\Omega_{N,x'},\delta)f(\Omega_{N,y'},\delta)f(\Omega_{N,z'}-\overline{\Omega},\varepsilon\delta).
\]
Here $x',y'$ and $z'$ are Cartesian coordinates with $z'$ axis being parallel to the magnetic field and nuclear polarization direction. The parameter $\varepsilon$ describes the suppression of longitudinal field fluctuations due to nuclear spin polarization and is given by
\begin{equation}
 \varepsilon = \sech \frac{B}{B_T}.
 \label{epsilon}
\end{equation} 

Since the maximum field $\Omega_0$, given by Eq.~\eqref{omega0}, is about $\sqrt{N}$ times larger than the characteristic fluctuation of hyperfine field in the equilibrium, Eq.~\eqref{delta}, even the small nuclear polarization degree $P=\overline{\Omega}/\Omega_0\ll 1$ strongly modifies the electron spin dynamics. This regime will be further addressed as ``additional field regime'', because the spin noise in this case is the same as in the presence of external field $\mathbf{\Omega}_B+\overline{\mathbf{\Omega}}$ and  $\varepsilon\approx 1$. In the opposite case of high nuclear spin polarization $P\lesssim 1$ the dispersion of $\Omega_{z'}$ drastically reduces since $\varepsilon\ll1$, and we call it ``fluctuations suppression regime''.

We note that the applicability of the model is limited by the condition $\varepsilon\gg1/\sqrt{N}$. From the physical point of view it ensures that the width of the distribution $\varepsilon\delta$ exceeds the characteristic correlation frequency of the hyperfine field $\sim\delta/\sqrt{N}$ (which for the sake of simplicity is assumed to exceed $1/T_2$, where $T_2$ is the dipolar relaxation time). From the formal mathematical considerations it can be obtained using the Berry–Esseen theorem for the rate at which the distribution converges to the normal one. Taking into account Eq.~\eqref{epsilon} one concludes that the applicability of the ``frozen'' nuclear polarization model is limited by condition $B<B_T\ln N$.

\section{Faraday geometry}
\label{sec:faraday}

When the external magnetic field and average nuclear field are oriented along the probe beam (Faraday geometry) the general equation~\eqref{general} for the spin noise spectrum can be recast as
\begin{multline}
\label{faraday}
 (s_z^2)_\omega=\frac{\sqrt{\pi}\tau_s}{\delta}\int_{-\infty}^\infty\d\Omega_z f(\Omega_z-\Omega_{\mathrm{tot}},\epsilon\delta) \\
 \times\int_0^\infty\d\Omega_\perp\Omega_\perp f(\Omega_\perp,\delta)
 \left\lbrace\cos^2\theta\frac{1}{1+\omega^2\tau_s^2}\right. \\
 \left.+\sin^2\theta\frac{1+(\omega^2+\Omega^2)\tau_s^2}{\left[1+(\omega-\Omega)^2\tau_s^2\right]\left[1+(\omega+\Omega)^2\tau_s^2\right]}\right\rbrace.
\end{multline} 
Here $\Omega_\perp=\sqrt{\Omega_x^2+\Omega_y^2}$ is the in-plane component of the Larmor frequency $\mathbf{\Omega}=\mathbf{\Omega}_N+\mathbf{\Omega}_B$, $\theta$ is the angle between $\mathbf{\Omega}$ and the $z$-axis, and $\Omega_{\mathrm{tot}}=\Omega_B+\overline{\Omega}$. In the additional field regime ($\overline{\Omega}\ll\Omega_0$) and for $\Omega_{\mathrm{tot}}\gg\delta$ (or $B\gg B_T/\sqrt{N}$) the integrals in Eq.~\eqref{faraday} can be solved analytically and we obtain
\begin{equation}
\label{faraday-add}
 (s_z^2)_\omega=\frac{\pi}{2}\left(1-\frac{\delta^2}{\Omega_{\mathrm{tot}}^2}\right)\Delta(\omega)+
 \frac{\pi\delta^2}{4\Omega_{\mathrm{tot}}^2}f(\omega-\Omega_{\mathrm{tot}},\delta).
\end{equation} 
This expression reduces to the trivial one in the fluctuations suppression regime ($\overline{\Omega}\lesssim \Omega_0$):
\begin{equation}
 (s_z^2)_\omega=\frac{\pi}{2}\Delta(\omega).
 \label{faraday-sup}
\end{equation} 
In this case the nuclei-induced precession related peak in the spectrum is negligible and the single zero-frequency Lorentzian peak is three times higher than in the equilibrium, cf. Eq.~\eqref{steady-limit}. The narrowing of the spin precession peak in Faraday geometry cannot be correctly described in the model of frozen nuclei, because the dynamics of transverse hyperfine field caused by the Knight field should be taken into account.

\begin{figure}[t]
 \includegraphics[width=0.49\textwidth]{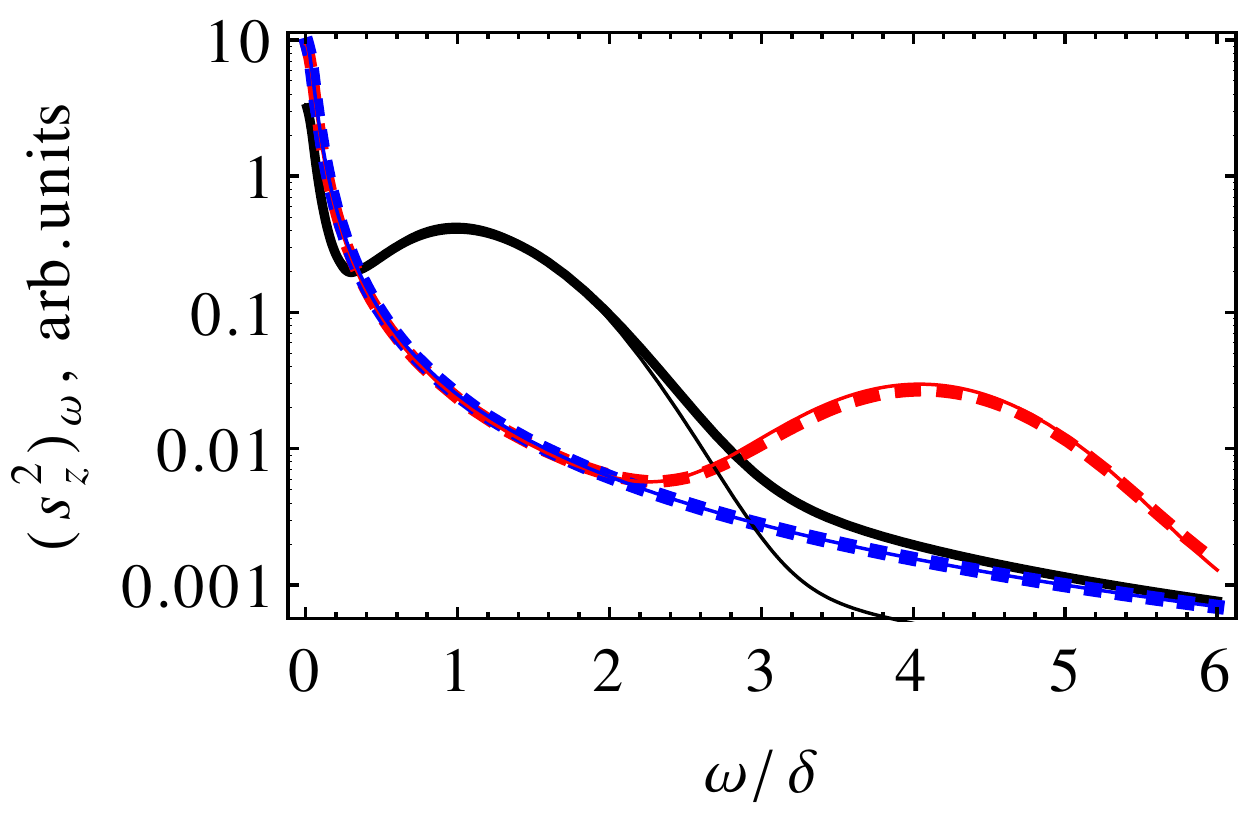}
\caption{Spin noise spectra of electrons in a longitudinal magnetic field. Thick curves are calculated after Eq.~\eqref{faraday} for the nuclear spin polarizations $P=0$ (black solid curve), $25\%$ (red dashed curve) and $90\%$ (blue dotted curve). The other parameters are $N=500$, $\tau_s\delta=20$ and $\Omega_B=\delta/10$. Thin curves are calculated after Eqs.~\eqref{steady-limit}, \eqref{faraday-add} and \eqref{faraday-sup} for the same parameters.}
\label{fig:faraday}
\end{figure}

Figure~\ref{fig:faraday} shows the electron spin noise spectra calculated after Eq.~\eqref{faraday} for different nuclear spin polarizations: thick solid curve corresponds to unpolarized nuclei, while thick dashed and thick dotted curves correspond to $P=25\%$ and $90\%$, respectively. In the absence of magnetic field and nuclear spin polarization, the spin noise spectrum in agreement with Ref.~\cite{NoiseGlazov} consists of a high and narrow zero-frequency peak and a low and broad peak centered at frequency $\delta$. With the increase of nuclear spin polarization the zero-frequency peak becomes three times higher, because the total field is three times more likely directed along $z$-axis. In the same time the spin precession peak shifts to higher frequency $\Omega_{\mathrm{tot}}$ and becomes symmetric but then rapidly disappears. The approximate expressions~\eqref{steady-limit},\eqref{faraday-add} and \eqref{faraday-sup} describe the spin noise spectra with the accuracy of about $10\%$, e.g. $\sim 1/\sqrt{N}$. Qualitatively the spin noise in the Faraday geometry can always be described by Eq.~\eqref{faraday-add} for the additional field regime.


\section{Voigt geometry}
\label{sec:voigt}

In the widely used Voigt configuration, where the magnetic field is applied in the plane transverse to the light propagation axis $z$, equation~\eqref{general} can be recast as
\begin{multline}
\label{voigt}
(s_z^2)_\omega=\frac{\tau_s}{2}\int\d\Omega_x\d\Omega_y\d\Omega_z f(\Omega_x-\Omega_{\mathrm{tot}},\varepsilon\delta)\times\\
 f(\Omega_y,\delta)f (\Omega_z,\delta)
 \times\left\lbrace\cos^2\theta\frac{1}{1+\omega^2\tau_s^2}+\right. \\
 \left.\sin^2\theta\frac{1+(\omega^2+\Omega^2)\tau_s^2}{\left[1+(\omega-\Omega)^2\tau_s^2\right]\left[1+(\omega+\Omega)^2\tau_s^2\right]}\right\rbrace,
\end{multline} 
where the notations are the same as in Eq.~\eqref{faraday} and the magnetic field is assumed to be applied along the $x$-axis. 

In additional field regime as in Sec.~\ref{sec:faraday} for $1/\sqrt{N}\ll P\ll1$ we arrive at
\begin{equation}
 \label{voight-add}
 (s_z^2)_\omega=\frac{\pi\delta^2}{4\Omega_{\mathrm{tot}}^2}\Delta(\omega)+\frac{\pi}{4}\left(1-\frac{\delta^2}{2\Omega_{\mathrm{tot}}^2}\right)f(\omega-\Omega_{\mathrm{tot}},\delta).
\end{equation} 

In the fluctuations suppression regime the zero-frequency peak becomes negligible because for $B\gtrsim B_T$ $\overline{\Omega}\sim\Omega_0\gg\delta$, while the spin precession peak dramatically narrows down and becomes higher. While the width of the peak exceeds spin relaxation rate $1/\tau_s$ the spectrum is given by
\begin{equation}
 \label{voight-sup}
 (s_z^2)_\omega=\frac{\pi}{4}f(\omega-\Omega_{\mathrm{tot}},\varepsilon\delta).
\end{equation} 
For nearly complete nuclear spin polarization $P\approx 1$ this condition is violated and the spectrum is described by the Voigt profile
\begin{equation}
 \label{voight-voigt}
 (s_z^2)_\omega=\frac{\pi}{4}V(\omega-\Omega_{\mathrm{tot}}; \varepsilon\delta, 1/\tau_s),
\end{equation} 
where
\[
 V(x;\sigma,\gamma)=\int_{-\infty}^\infty\frac{e^{-x'^2/\sigma^2}}{\sigma\sqrt{\pi}}\frac{\gamma\d x'}{\pi\left[(x-x')^2+\gamma^2\right]},
\]
is the convolution of the Gaussian and Lorentzian functions. In the limiting case of $\varepsilon\delta\ll1/\tau_s$ one arrives at usual Lorentzian peak at the Larmor precession frequency
\begin{equation}
 (s_z^2)_\omega=\frac{\pi}{4}\Delta(\omega-\Omega_{\mathrm{tot}}).
 \label{voight-lorentz}
\end{equation}
Hence the evolution of the spin noise spectrum in the transverse magnetic field reflects the suppression of nuclear spin fluctuations.

\begin{figure}[t]
 \includegraphics[width=0.48\textwidth]{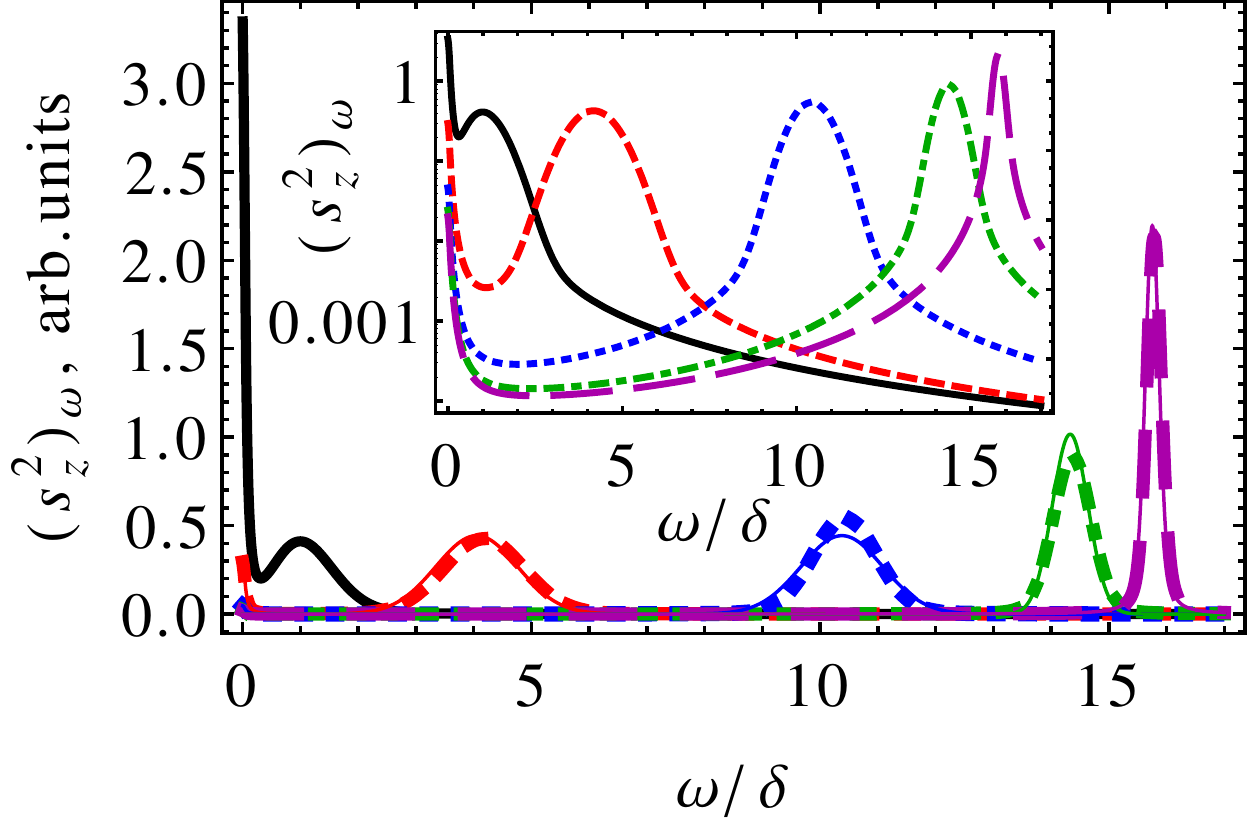}
 \caption{Spin noise spectra in a transverse magnetic field. Thick curves are calculated after Eq.~\eqref{voigt} for nuclear polarizations $P=0$ (solid black curve), $25\%$ (dashed red curve), $65\%$ (dotted blue curve), $90\%$ (dash-dotted green curve) and $99\%$ (long-dashed magenta curve) with the same parameters as in Fig.~\ref{fig:faraday}. The thin curves corresponding to $P=25\%$ and $65\%$ are calculated after Eq.~\eqref{voight-add}, and the curves corresponding to $P=90\%$ and $99\%$ are calculated after Eq.~\eqref{voight-sup} and~\eqref{voight-voigt} respectively. The inset shows the thick curves in the logarithmic scale.}
\label{fig:voigt}
\end{figure}

Figure~\ref{fig:voigt} presents the electron spin noise spectra calculated for the Voigt geometry and several values of the external magnetic field. Figure shows that the precession-induced peak shifts upon increasing the nuclear spin polarization along the $x$-axis and follows the total Larmor frequency $\Omega_{\mathrm{tot}}$ while the low-frequency peak decreases abruptly. For polarization $P\lesssim1$ the fluctuations of the nuclear field get suppressed and the peak dramatically narrows down and becomes higher up to $\sqrt{N}$ times. The transformation of the spectrum can by described by analytical expressions~\eqref{voight-add}-\eqref{voight-lorentz} with the accuracy scaling as $1/\sqrt{N}$.

\section{Nuclear spins $I>1/2$}
\label{sec:I}

\begin{figure}[bthp]
 \includegraphics[width=\linewidth]{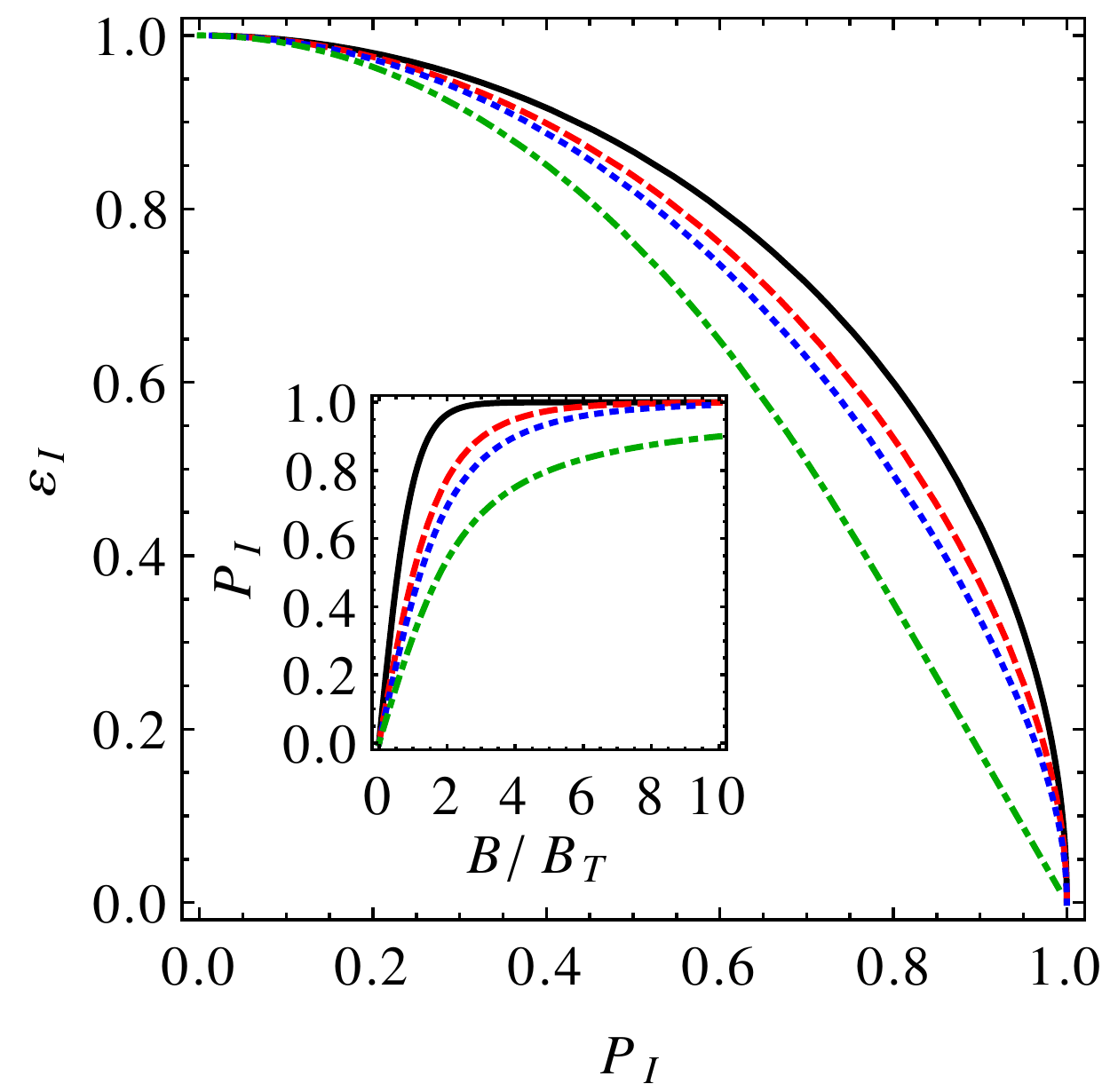}
 \caption{Degree of nuclear fluctuations suppression as a function of $P_I$ for $I=1/2$ (black solid curve), $3/2$ (red dashed curve), $5/2$ (blue dotted curve) and $I\to\infty$ (green dash-dotted curve). The inset shows nuclear spin polarization degree as a function of $B/B_T$ for the same $I$.}
\label{fig:Pepsilon}
\end{figure}

In the above analysis for the sake of simplicity we assumed nuclear spins to be $I=1/2$. In the general case of $I>1/2$ one has to revise Eqs.~\eqref{P} and~\eqref{epsilon} for average Larmor frequency and for the decrease of dispersion. The spin polarization $P_I$ is given by the Brillouin function $P_I=\mathcal B_I(B/B_T)$, where $B_T=kT_N/(I\hbar\gamma)$~\cite{kittel2004wiley}. The reduction of the dispersion can be described by parameter $\varepsilon_I$, which is defined by
\begin{equation}
\varepsilon_I=\sqrt{\frac{\left\langle (I_{z'}-P_I I)^2\right\rangle_B}{\left\langle I_{z'}^2\right\rangle_0}}.
\label{epsilonI}
\end{equation} 
Here the angular brackets, as above, stand for the averaging over the time and/or the ensemble and the subscript denotes magnitude of the external magnetic field, $B_T$ is assumed to be fixed. In general, the parameter $\varepsilon_I$ can be expressed as~\cite{ll5_eng}
\begin{equation}
 \varepsilon_I=\sqrt{\frac{3I}{I+1}B_T\frac{\partial P}{\partial B}}.
\end{equation} 
In the classical limit $I\gg 1$ the Brillouin function transfers to the Langevin function, which allows one to obtain simple analytical expressions
\begin{subequations}
\begin{equation}
 \label{P-inf}
 P_\infty=\coth\frac{B}{B_T}-\frac{B_T}{B},
\end{equation} 
\begin{equation}
\label{e-inf}
  \varepsilon_\infty=\sqrt{3}\sqrt{\frac{B_T^2}{B^2}-{\sinh}^{-2}\frac{B}{B_T}}.
\end{equation} 
\end{subequations}

In addition to the described effects in case of $I>1/2$ the magnetic field suppresses the transverse components of the hyperfine field, $\Omega_{x',y'}$ (or $I_{x',y'}$). One can show that
\[
 \frac{\left\langle I_{x',y'}^2\right\rangle_B}{\left\langle I_{x',y'}^2\right\rangle_0}=\frac{3}{2}-\frac{3IP_I^2}{2(I+1)}-\frac{\varepsilon_I^2}{2},
\]
where the notations are the same as in Eq.~\eqref{epsilonI}.

In the inset of Fig.~\ref{fig:Pepsilon} the nuclear spin polarization degree is plotted as a function of $B/B_T$ for different nuclear spins $I$. The larger $I$ the flatter is the dependence of $P_I(B)$. Ultimately the shape of the spin noise spectra is defined by the relation between $P_I$ and $\varepsilon_I$, presented in Fig.~\ref{fig:Pepsilon}. This dependence is similar for $I=1/2,3/2$ and $5/2$. On the whole one can see that the suppression of spin fluctuations due to nuclear polarization is more efficient for large $I$.

\section{Conclusions}
\label{sec:conclude}

In this paper, we have developed a microscopic theory of spin noise of localized electrons interacting with cooled nuclear spin bath. Even relatively weak magnetic fields of about several Gauss effectively polarize nuclear spins and drastically modify spin noise spectra. Magnetic field applied in the Faraday geometry leads to the three-fold amplification of zero-frequency peak in the spin noise spectrum and complete suppression of spin precession peak. On the contrary, the application of magnetic field in the Voigt geometry suppresses the zero frequency peak and shifts the other one to the higher frequency range. Importantly the substantial nuclear polarization degree $P\lesssim 1$ suppresses the fluctuations of the hyperfine field and the precession-related peak becomes $\sim\sqrt{N}$ times narrower and higher.

The majority of the described effects can be studied by a conventional spin-noise technique.
However the suppression of hyperfine field fluctuations takes place, when the Larmor precession frequency $\overline{\Omega}$ is comparable with $\Omega_N\sqrt{N}$, which typically has the order of several GHz (corresponds to magnetic fields $\sim 0.1$ T). So it can be easier to observe this effect in semiconductors with low nuclear spin concentration, for example, in ZnSe doped with Fluorine~\cite{greilich2012ZnSe:F} or ZnO doped with Aluminum~\cite{ZnOnoise}, or in singly charged quantum dots made of the same materials~\cite{Liu2007,norris2008doped}. In the latter case each electron interacts with $N\sim100$ nuclear spins making it possible to study complete nuclear spin polarization
in the hundreds of megahertz frequency range. 
Moreover, for low nuclear spin concentration, the interaction between nuclear spins is suppressed which facilitates dynamical nuclear polarization~\cite{Hubert2004,dyakonov1975}.
Alternatively a pulsed laser light source~\cite{Mueller2010,Gigahertz2010} or ultrafast spin-noise spectroscopy~\cite{Oestreich-review, starosielec:051116, Berski-fast-SNS} can be used for such studies in GaAs-type semiconductors.

To conclude, we propose to use the spin noise spectroscopy as a tool to investigate nuclear spin temperature and nuclear spin dynamics. The developed theory allows one to extract nuclear spin temperature from the electron spin noise spectra in the presence of magnetic field and, in principle, to study the slow relaxation of nuclear spin temperature by measuring electron spin noise spectra as a function of time.

\section*{Acknowledgements}

The author gratefully acknowledge A. Greilich, M. M. Glazov and E. L. Ivchenko for helpful discussions and careful reading of the manuscript.
This work was partially supported by 
the Russian Foundation for Basic Research, 
Dynasty Foundation 
and St.-Petersburg Government Grant.


%

\end{document}